\documentclass[
twocolumn, prc,floats,preprintnumbers,amsmath,amssymb]{revtex4}
\usepackage{epsfig}
\usepackage{amsmath}
\renewcommand{\(}{\left(}
\renewcommand{\)}{\right )}
\renewcommand{\[}{\left [}
\renewcommand{\]}{\right ]}

\def\pa{\partial}

\def\bea{\arraycolsep .1em \begin{eqnarray}}
\def\eea{\end{eqnarray}}

\let\be=\beta
\let\no=\nonumber

\def\eq#1{(\ref{#1})}
\def\refr#1{\cite{#1}}

\def\s0#1#2{\mbox{\small{$ \frac{#1}{#2} $}}}
\def\0#1#2{\frac{#1}{#2}}

\def\cpl#1#2#3{Chin. \ Phys.\ Lett. \ {\bf #1}, #2 (#3).}
\def\ctp#1#2#3{Commun.\ Theor.\ Phys. \ {\bf #1}, #2 (#3).}

\def\npa#1#2#3{Nucl. Phys. {\bf A #1}, #2 (#3).}

\def\pra#1#2#3{Phys. Rev.  {\bf A #1}, #2 (#3).}

\def\prc#1#2#3{Phys. Rev.  {\bf C #1}, #2 (#3).}

\def\prl#1#2#3{Phys. Rev. Lett. {\bf #1}, #2 (#3).}

\def\prep#1#2#3{Phys.\ Rep.\ {\bf #1}, #2 (#3).}

\def\rmp#1#2#3{Rev.\ Mod.\ Phys.\ {\bf #1}, #2 (#3).}

\begin{document}
\title{Determination of Landau Fermi-liquid parameters of strongly
interacting fermions by means of a nonlinear scaling transformation}
\author{Ji-sheng Chen\footnote{chenjs@iopp.ccnu.edu.cn}}
\address{Physics Department \& Institute of Particle Physics,
Central China Normal University, Wuhan 430079, People's Republic of
China}
\begin{abstract}
A nonlinear transformation approach is formulated for the correlated
    fermions' thermodynamics through a medium-scaling effective action.
An auxiliary implicit variable-effective chemical potential is
    introduced to characterize the non-Gaussian fluctuations physics.
By incorporating the nonlocal correlation effects, the achieved
    grand partition function is made of coupled highly nonlinear
    parametric equations.
Analytically, the low temperature expansions
    for the strongly interacting unitary Fermi gas are performed for the
    adiabatic compressibility-sound speed and specific heat with the
    Sommerfeld lemma.
The expressions for the Landau Fermi-liquid parameters $F_0^s$ and
    $F_1^s$ of the strongly interacting fermion system are obtained.
As a universal constant, the effective fermion mass ratio is
    $m^*/m=\frac{10}{9}$ at unitarity.

\end{abstract}



\maketitle

The Landau theory of Fermi liquid  plays the key role in
    understanding the novel interacting fermions quantum many-body
    physics\refr{Landau1956}.
Either in the non-relativistic or relativistic frameworks, the
    central task is how to calculate the Landau Fermi-liquid parameters\refr{Baym1991,Baym1976,Friman1999,Brown2002}, which
    contributes to addressing the bulk or kinetic transport
    properties in terms of the quasi-particle
    viewpoint.
The conventional techniques for calculating the effective fermion
    mass and/or ground state energy will involve the various coupled
    integral equations for the multi-points correlation Green functions
    exhibited by the Galitskii, Bethe-Sepeter or Bethe-Goldstone
    expansions\refr{walecka1971}.

In recent years, considerable efforts for understanding the
    physics from Bardeen-Cooper-Schrieffer to Bose-Einstein
    condensation(BCS-BEC crossover) with
    ultra-cold atomic Fermi gases have been made.
The two-body interaction strength can be tuned with the Feshbach
    resonance.
At unitarity, the divergent scattering length with a zero energy
bound
    state can manifest the universal properties\refr{Heiselberg2000,Ho2004,Giorgini2007,Lee2008}.

The fundamental
    issue on the unitary fermions thermodynamics is on the symmetric fermions ground
    state energy\refr{Heiselberg2000,Ho2004}.
Another exciting one concerns the asymmetric system
    thermodynamics.
To describe or interpret the phase separation properties achieved by
the experimentally trapped systems,
    the Landau effective fermion mass
    becomes the key dynamical parameter\refr{Lobo2006,Pilati2008,Yong2008}.
Like in the  ground state energy, the infinite scattering
    length will certainly drop out in the effective mass expression, which excites considerable attempts\refr{Lobo2006,Pilati2008,Combescot2007,Baker1999}.
The existing values are in the regime ${m^*}/{m}\sim
    1.04-2.5$. What is the exact value of ${m^*}/{m}$?

In this Letter, we will develop a non-perturbative statistical
mechanical method
    to understand the novel strongly correlating physics.
The comprehensive thermodynamical quantities can be fixed by the
    underlying partition function.
For instance,
    the entropy incorporating the dynamical and quantum correlations can be
    derived from the grand thermodynamical potential.
As we will see,
    the entropy reasonably characterizing the strong
    correlation information is crucial in determining the Landau
    parameters while ensuring the strict thermodynamical
    self-consistency.

The basic prospective is that the single particle
    spectrum will be modified by the collective dispersive effects.
The medium-dependent effective action
    allows a natural implementation of the collective effects
    as discussed in nuclear many-body literature\refr{Friman1999,Brown2002}.
One can imagine that the bare interaction potential is
    renormalized by the surrounding environment and the
    single particle spectrum will be modified.
At unitarity, the medium dependence of the interaction strength can
    cancel the infrared divergences and lead to a finite physical
    result.
Fixing the Landau Fermi-liquid parameters of strongly interacting
    fermions at unitarity is the motivation.

In order to account for the unusual fluctuation/correlation effects,
    we work with the medium-scaling Hamiltonian in
terms of the grand canonical ensemble framework\refr{chen2007} \bea
\label{Hamiltonian}
    \tilde{H}=&&-\int d^3x
    \psi_\alpha ^*(x) (\0{\nabla ^2}{2m}-\mu_{r}[n,T])\psi _\alpha (x)\no\\
    &&+\0{U_{\mbox{eff}}^*[n,T]}2 \int d^3x\psi_\alpha ^*(x)\psi^*_\beta (x)
    \psi_\beta(x)\psi_\alpha (x).
\eea In equation \eq{Hamiltonian},
    $\alpha,\be=\uparrow (a), \downarrow (b)$
    represent the hyperfine-spin projection Ising-variables while the $m$ is the bare fermion mass.
The bare coupling constant $U_0={4 \pi a}/{m}$  has
    been substituted by a medium dependent functional through a
    specific transformation
\bea \label{poten}
    U_{\mbox{eff}}^*[n,T]=\0{U_0}{1-\012 m_D^2 U_0},~~~~~
    m_D^2=\(\0{\pa n}{\pa \mu^*}\)_T.
\eea The $n$ is the particle number density and $\mu^*$ the
    effective chemical potential defined below through equation \eq{final2}.

The \textit{minus} sign in the denominator of equation \eq{poten}
incorporates the
    alternating frustration function of the surrounding environment.
The frustration spirit is consistent with the
    general Le Chatelier's stability principle in thermodynamics\refr{Landau,chen2007}.
The nonlocal correlation physics characterized by the
    alternating minus sign coincides with the second law of thermodynamics.
Furthermore,
    the many-body correlation effects are taken into account as a
    spontaneously generated \textit{single-body} potential $\propto \mu_r[n,T]$.
Sticking to the
    medium dependence of $U_{\mbox{eff}}^*$,
    the additional compensatory term $\delta {\cal H}\propto \mu_{r}[n,T]N_\alpha$
   can enforce the energy-momentum
    conservation law.

Two steps will be taken to give the grand
    thermodynamical potential with equation \eq{Hamiltonian} and the medium-scaling interaction, equation \eq{poten}.
Firstly,
    the \textit{shifted} relative minimum $\tilde{\Omega} (T,{\tilde \mu }) $ is
    obtained by fixing the interaction analogous to the Hartree approximation with the linear bare potential.
Secondly, the absolute minimum ${\Omega} (T,{\mu }) $ for the given
    chemical potential $\mu $ is derived with the constraint Legendre
    transformation correspondence relations between $\mu $ and $N$ or
    $T$ and $S$\refr{chen2007,chen2007-1}.

The achieved equation of state/grand thermodynamical potential can
be presented as the coupled \textit{parametric equations} formalism
\bea\label{final}
    P&&=P_{\mbox{ideal}}(T,\mu^*)+\0{\pi
    a_{\mbox{eff}}}{m}n^2 +{\cal C}
    \(\0{2\pi a_{\mbox{eff}}}{m}\)^2n^3,\\
    \label{final2}
    \mu &&=\mu^*+\0{2\pi a_{\mbox{eff}}}{m}n +{\cal C}
    \(\0{2\pi a_{\mbox{eff}}}{m}\)^2n^2,
     \eea
with \bea
    P_{\mbox{ideal}}=\0{2T}{\lambda^3}f_{5/2}(z'),
\eea being a function of the effective chemical potential. The
employed notation $a_{\mbox{eff}}$ is defined by
$U^*_{\mbox{eff}}\equiv 4\pi
    a_{\mbox{eff}}/m$.

With the implicit collective variable $\mu^*$ introduced by the
single particle Green function self-consistent equation \eq{final2},
    the total number density
    $n=n_\uparrow+n_\downarrow=2n_\uparrow$ is
\bea \label{density} n(T,\mu^*)\equiv 2\int_k
    f_k =\0{2}{\lambda^3}f_{3/2}(z'),
\eea with the momentum integral symbol $\int_k=\int d^3{\bf
k}/(2\pi)^3$ and quasi-particle Fermi-Dirac distribution functions
defined as\bea
    \label{dirac} f_k&&=\01{z'^{-1}e^{\be \0{{\bf
    k}^2}{2m}}+1},~~~
    z'=e^{\be\mu^*}.
\eea The $\lambda =\sqrt{2\pi/(mT) }$ is the mean thermodynamical de
    Broglie  wavelength. The $\be$ is the inverse temperature (with $k_B=\hbar=1$ throughout the letter).
The {\em effective fugacity} $z'$ is analogous to fugacity
     $z=e^{\be\mu}$.

The first two terms of equation \eq{final} have
    exactly the same structure of the mean-field theory with fixed local interaction, which is linear-like.
The last correction term $\propto {n^3}$ is picked up
    in a thermodynamical way by relaxing the medium-dependence of
    interaction potential;
it describes the non-Gaussian correlation physics and
    of non-local characteristic.
They are the \textit{rearrangement}
    effects of single particle energy spectrum shift reflected by the
    Brueckner-Bethe-Goldstone (BBG) technique with medium-dependent
    interaction potential\refr{Brown2002}.
From the statistical field theory viewpoint, this term guarantees
the energy-momentum conservation law.

Associated with $U^*_{\mbox{eff}}[n,T]$,
    the shift strength ${\cal C}$ can be identified to be
\bea
    \label{c}{\cal C}(T,\mu^*) &&=\012\(\0{\pa m_D^2}{\pa n}\)_T\no\\
    &&=\01{2T}
    \0{f_{-1/2}(z')}{f_{1/2}(z')}.
\eea Furthermore, the entropy density $s=S/V$ can be derived with
the Hamiltonian-Jacobi
    \textit{implicit variable} method from  the coupled equations, \eq{final} and \eq{final2}
\bea s&&=\(\0{\pa P}{\pa
    T}\)_\mu\no\\
    &&=s_{\mbox{ideal}}+{\cal D}(\0{2\pi
    a_{\mbox{eff}}}{m})^2n^2,
\\
     s_{\mbox{ideal}}&&\equiv
     -2\int _k \[f_k\ln f_k+(1-f_k)\ln (1-f_k)\],
\\
\label{d}
    {\cal D}(T,\mu^*)&&=-\012\(\0{\pa m_D^2}{\pa T}\)_n.
\eea

There is a dual  relation between ${\cal C}$ and ${\cal D}$,
    which can be seen from equations \eq{c} and \eq{d}.
It is worthy noting that the correlation factor ${\cal
    D}(T,\mu^*)$ explicitly
    characterizes  the temperature fluctuation in addition to density
    fluctuation.
For the
    non-relativistic scenario, the factor ${\cal D}$
can be furthermore simplified \bea
    {\cal D}(T,\mu^*)=-\0{m_D^2}{4T}+\0{3{\cal C}}{2T}n.
\eea

Correspondingly, the entropy  and energy density
    $\epsilon=E/V=-P+\mu n +Ts$ can be also reduced to the compact formalisms
    with the explicit $\propto n^2, n^3$ correction
    terms
\bea\label{entropy} s&&=s_{\mbox{ideal}}-\0{m_D^2}{T} \(\0{\pi
    a_{\mbox{eff}}}{m}\)^2 n^2 +\0{3{\cal C} }{2T} \(\0{2\pi
    a_{\mbox{eff}}}{m}\)^2n^3;
\\\label{final3}
    \epsilon&&=\epsilon_{\mbox{ideal}}+\032\(1-\01{3(1-\0{2\pi m_D^2
    a}{m})}\)\0{\pi a_{\mbox{eff}}}{m} n^2\no\\&& ~~~~~+\0{3{\cal C}
    }{2} \(\0{2\pi a_{\mbox{eff}}}{m}\)^2n^3.
\eea
 With $P_{\mbox{ideal} }=\023 \epsilon
_{\mbox{ideal}}$,
    the Virial Theorem $PV=\023 E$ of \textit{ideal} gas is satisfied at unitarity\refr{Ho2004},
    which is obvious by comparing equation \eq{final} and equation \eq{final3}.
The vanishing $\propto n^2,n^3$ terms
    in the entropy density at $T\rightarrow0$ ensures the third law of thermodynamics as expected.

The thermodynamical properties can be characterized by the coupled
    parametric equations \eq{final} and \eq{final2} or equations \eq{entropy} and
    \eq{final3}.
Relevant formulae can be expressed explicitly in terms of $s=S/V$
    etc.
    and according to $T$ and the implicit variable $\mu^*$.

For example,
    the sound speed squared is given by the adiabatic
    compressibility $\kappa_S=\({\pa n}/{\pa P}\)_S$ according to
\bea
    c^2&&=\(\0{\pa P}{\pa m n}\)_S=\01m
    \[\(\0{\pa P}{\pa n}\)_s+\0sn \(\0{\pa P}{\pa s}\)_n\].
\eea The specific heat per particle at constant volume  is \bea
    \0{C_V}{N}=\0{T}{n}\(\0{\pa s}{\pa T}\)_n.
\eea The specific heat $C_P$ per particle at constant pressure  is
    calculated according to
\bea \0{C_P}{N}=\0Tn\[\(\0{\pa s}{\pa
    T}\)_P-\0sn \(\0{\pa n}{\pa T}\)_P\].
\eea The general expressions for $c^2$, $C_V$, $C_P$ etc. at
    finite temperature involve the high order susceptibilities described by such as
    $f_{-5/2}(z')$.
They are quite lengthy and are emitted for
    brevity.
Obviously, the important role has been played by the entropy
    incorporating the dynamical high order and quantum
    fluctuation/correlation effects.

Now we take the Sommerfeld lemma to do the low temperature expansion
    in order to give the Landau parameters.
To characterize the strong correlation physics,
    the expansion is in terms of $T/\mu^*=({\ln z'})^{-1}$ instead of
    directly according to $({\ln z})^{-1}$.

Let us start from the standard Fermi integral
$f_{5/2}(z')$\refr{walecka1971} \bea \label{f52}
    f_{5/2}(z' )&&=\0{4 }{3
    \sqrt{\pi}}I(\alpha),~~~
    \alpha =\0{\mu^*}T;~~~~~
    I(\alpha)\equiv\int _{-\alpha} ^\infty
    \0{(x+\alpha)^{3/2}}{e^x+1}dx.\no
\eea In the strong degenerate regime with $\mu^*\gg T$, $I(\alpha)$
    and Fermi integral $f_{5/2}(z')$ can be approximated by
\bea
    I(\alpha)&&=\frac{2 {\alpha }^{5/2}}{5}+ \frac{\pi^2
    {\alpha }^{1/2}}{4} - \frac{7 \pi^4}{960 {\alpha
    }^{3/2}}+
 \cdots;
\\\label{expansion}
    f_{5/2}(z')&&= \frac{8 {\mu^*}^{5/2}}{15 \sqrt{\pi } T^{5/2}}+
      \frac{\pi ^{3/2} {\mu^*}^{1/2}}{3 T^{1/2}}  -
    \frac{7 \pi ^{7/2} T^{3/2}}{720 {\mu^*}^{3/2}}
+\cdots. \eea
 From equation \eq{expansion},
    the low temperature expansions of the remaining Fermi integrals can be obtained by using
\bea
    f_{j-1}(z')=z'\0{\pa f_j(z')}{\pa z'}=T \0{\pa f_j(z')}{\pa \mu^*}.
\eea

With \bea
    f_{3/2}(z')=\frac{4 {\mu^*}^{3/2}}{3 \sqrt{\pi } T^{3/2}}+
  \frac{\pi ^{3/2} T^{1/2}}{6 {\mu^*}^{1/2}}+
\frac{7 \pi ^{7/2} T^{5/2}}{480 {\mu^*}^{5/2}} +\cdots, \eea and the
particle number density $ n=k^3_f/{(3 \pi }^2)$, the effective
    chemical potential can be inversely expanded in terms of the ``Fermi"
    momentum $k_f$
\bea
    \label{chem-low}\mu^*= \frac{k_f^2}{2 m} - \frac{
    \pi^2 mT^2}{6 k_f^2} - \frac{ {\pi
    }^4 m^3T^4}{10 k_f^6}
    +\cdots.
\eea

Inserting the intermediate result, equation \eq{chem-low}, into the
    expansions such as equation \eq{expansion} of the various Fermi integrals
    $f_j(z')$ ($j=-\052,-\032,\cdots,\052$) in $C_V$, $c^2$ and $P$ etc.,
    the thermodynamical
    quantities can be expanded according to $T/T_f$.
Here, $T_f=k_f^2/(2m)$ is the ``Fermi" characteristic temperature.
The calculation is quite lengthy by not complicated and the
    explicit expressions can be obtained.

For example,
    the expansion for pressure is
\bea
    \label{press-ex} P&&=\(1 +
    \frac{5 a  k_f}{3 \left( \pi  -2 a  k_f  \right) }+ \frac{10 (ak_f)^2}{9
    {\left( \pi-2 a  k_f \right ) }^2}\) \frac{k_f^5}{15 m {\pi
    }^2}
\no\\
    &&+\(1 - \frac{8(a k_f)^3}{9 {\left( \pi -2 a  k_f \right) }^3}\)\frac{k_f m T^2}{9}\no\\
    &&-\(1  -
    \frac{176(a k_f)^2}{81 {\left( \pi -2 a  k_f \right) }^2}+
    \frac{236(a k_f)^3}{81 {\left( \pi -2 a  k_f \right) }^3}\right. \no\\
    && ~~~~~~~~\left.- \frac{40(a k_f)^4}{27 {\left( \pi -2 a  k_f
    \right) }^4} \)\frac{\pi ^2 m^3T^4}{15 k_f^3}+\cdots,
\eea
 where the factors not in the brackets are those
for the ideal
    Fermi gas.
By keeping up to the lowest order, equation \eq{press-ex} gives the
    universal  coefficient $\xi=\049$ to which much attention has been paid.
At unitarity, the universal coefficient of the next order is
    $\0{10}9$.
It coincides with the first order coefficient of $C_V/N$, equation
\eq{mass}, when $|a|=\infty$. However, the general expressions are
quite different from each other.
\begin{figure}[htb]
  \centering
   \includegraphics[width = .45\textwidth]{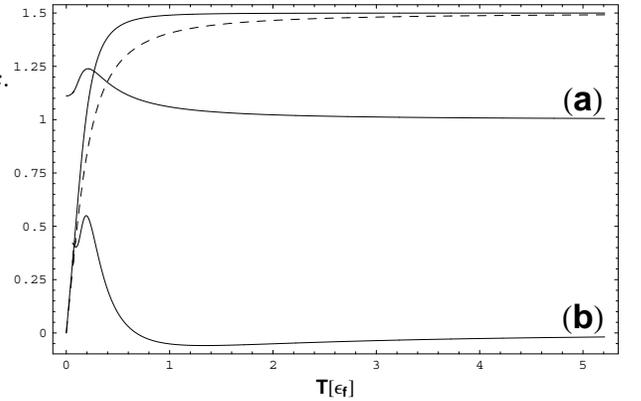}
   \caption{\small
    Universal specific heat: The above solid curve is $C_V/N$
    of the unitary gas while the dashed one  is $C_V^0/N$ for
    ideal Fermi gas. The labeled (a) curve is the $C_V/C_V^{(0)}$ ratio
    while the indicated (b) is for $\pa^2 \mu /\pa n\pa T$.}\label{fig1}
\end{figure}

The sound speed squared is expanded as \bea
    c^2
    =&&\(\0{1+F_0^s}{1+\013 F_1^s}\) \frac{k_f^2}{3 m^2}+\(1  -
  \frac{16 a  k_f}{5 \left( \pi  -2 a  k_f \right) }\right.\no\\
  &&\left. -
  \frac{16(ak_f)^2}{5 {\left( \pi  -2 a  k_f \right) }^2} -
  \frac{32(ak_f)^3}{45 {\left( \pi  -2 a  k_f \right) }^3} + \frac{16(ak_f)^4}{45 {\left( \pi  -2 a  k_f \right) }^4}\right.\no\\
  &&\left. +
  \frac{16 \left( -10(ak_f)^2 + 9 a  k_f \pi  \right) }
   {5 \left( 40(ak_f)^2 - 36 a  k_f \pi  + 9 \pi^2 \right) }\)\frac{5 \pi^2 T^2}{9 k_f^2}+\cdots.
\eea The lengthy second expansion coefficient is still $\0{10}9$ at
unitarity. By comparing with those of Landau Fermi-liquid
theory\refr{Baym1991},
    the first expansion coefficient is
\bea
    \0{1+F_0^s}{1+\013 F_1^s}=1
    + \frac{2 a k_f}{\pi -2 a k_f}+
  \frac{20(ak_f)^2}{9 {\left(\pi -2 a  k_f  \right) }^2} + \frac{8(ak_f)^3}{9 {\left(\pi -2 a  k_f  \right) }^3},\no\\
\eea

The lowest order of the low temperature expansion for specific heat
gives the effective mass definition \bea
    \0{C_V}{N}&&=\(\0{m^*}m\)\frac{\pi^2 m T}{k_f^2}-\(1  -
  \frac{88(ak_f)^2}{27 {\left( \pi -2 a  k_f  \right)
  }^2}\right. \no\\
    &&~~~\left. + \frac{40(ak_f)^3}{27 {\left( \pi-2 a  k_f   \right)
    }^3}\)\frac{6  {\pi
      }^4 (mT)^3}{5 k_f^6}+\cdots;\\
\0{C_P}{N}&&=\(\0{m^*}m\)\frac{\pi^2 mT}{k_f^2}
    +\(1 - \frac{160 a k_f}{3 \pi-4 a  k_f }\right. \no\\
    &&\left.
  +\frac{320 a  k_f}{9 \left( \pi -2 a  k_f  \) }
  +\frac{64(ak_f)^2}{9 {\left( \pi -2 a  k_f  \right) }^2}
  - \frac{40(ak_f)^3}{9 {\left( \pi -2 a  k_f  \right) }^3} \right.\no\\
  &&\left.
 -\frac{20 \left( -8(ak_f)^2 + 3 a  k_f \pi  \right) }
   {9 \left( 8(ak_f)^2 - 8 a  k_f \pi  + 3 \pi^2 \right) }\)\frac{2  \pi^4 (mT)^3}{15
   k_f^6}+\cdots;\no\\
   \\\label{mass}
   \0{m^*}m&&=1+\013
    F_1^s=1 + \frac{4(ak_f)^2}{9 {\left( \pi -2 a
    k_f \right) }^2}.
\eea The explicit difference between $C_V/N$ and $C_P/N$
    is from the second order.
At unitarity with $|a|=\infty$, the effective mass ratio is $
{m^*}/m=\0{10}9$, which agrees with the
    previous attempts $1.04-1.17$ \refr{Lobo2006,Pilati2008,Combescot2007}.
The effective fermion mass is the dynamical
    ``higher order" effect because the linear term $\propto k_fa$ is absent in the
    weak coupling expansion\refr{walecka1971}.
The reciprocal term $\propto 1/(k_fa) $ does appear
    in the strong interaction regime.

The universal property of unitary Fermi gas can be also explored by
such as the specific heat at unitarity. The scaling function
$C_V/C_V^0 $ for
     $C_V$  to $C_V^0$ of ideal Fermi gas is $\0{10}9$
    at $T\rightarrow 0$;
    it must be saturated as $C_V/C_V^0\rightarrow 1$ in the Boltzmann
    limit.
It is not a naive monotonously decreasing function with
    the increase of the rescaled temperature $T/T_f$.
The numerical study indicates that this ratio
    has a knee peak at $T^*\approx 0.21 T_f$ although $C_V/N$ itself is a monotonously increasing
function (different from the finite interaction scenario). This can
be attributed to the competition between the particle
    number and temperature fluctuations.
To characterize this unusual behavior, the mixing
    susceptibility $\pa ^2 \mu /\pa T\pa n$ is shown in
    Fig.\ref{fig1}, where a similar knee peak around $T^*$ appears.

To conclude, the medium scaling spirit is
    realized by a nonlinear transformation for the effective interaction and consequently for the physical chemical
    potential.
The introduced auxiliary implicit variable-effective chemical
    potential characterizes the correlation physics and makes the exact
    grand partition function appear as the highly nonlinear coupled parametric-equations.

The low temperature expansion in terms of $(\ln z')^{-1}$
(consequently eliminating $z'$) can be performed analytically with
    the Sommerfeld method.
The effective fermion mass and sound speed squared have been
    explicitly calculated.
As the important dynamical parameter in Landau Fermi-liquid theory,
    the calibrated effective mass is reasonably consistent with some Monte-Carlo attempts.
The parametric-equations formulation can gauge the
    infrared singularity manifested by the low-energy long wavelength thermodynamics
    of the three-dimensional contact interaction fermions system.

\acknowledgments{The author thanks Professor J.-R Li and Dr. X.-j
Xia for the beneficial discussions. Supported by the Natural Science
Foundation of China under Grant No. 10875050, 10675052 and MOE of
China
    under projects No.IRT0624.}

\end{document}